\documentclass[prd,reprint,nofootinbib,superscriptaddress,preprintnumbers,showpacs]{revtex4-1}
\usepackage{amsfonts}
\usepackage{amssymb}
\usepackage{amsmath}
\usepackage{graphicx,color}
\usepackage{dcolumn}
\usepackage{epsfig}
\usepackage{bm}
\usepackage{bbm}
\usepackage{ulem}
\usepackage{hyperref}
\usepackage{lineno}
\usepackage{float}
\usepackage{subfigure}
\usepackage{slashed}
\usepackage{CJKutf8}
\allowdisplaybreaks[4]

\begin{document}
	
	\title{Single spin asymmetry $ A _ { U L } ^ { \sin ( 2 \phi _ { h } ) }$ in dihadron production in SIDIS}
	\author{Ren Yang}
	\affiliation{School of Physics and Optoelectronics Engineering, Anhui University, Hefei 230601, People's Republic of China }
	\author{Yangyang Yu}
	\affiliation{School of Physics and Optoelectronics Engineering, Anhui University, Hefei 230601, People's Republic of China }
	\author{Qihang Zhou}
	\affiliation{School of Physics and Optoelectronics Engineering, Anhui University, Hefei 230601, People's Republic of China }
	\author{Gang Li}
	\affiliation{School of Physics and Optoelectronics Engineering, Anhui University, Hefei 230601, People's Republic of China }
	\author{Mao Song}
	\affiliation{School of Physics and Optoelectronics Engineering, Anhui University, Hefei 230601, People's Republic of China }
	\author{Xuan Luo}	
	\email{xuanluo@ahu.edu.cn}
	\affiliation{School of Physics and Optoelectronics Engineering, Anhui University, Hefei 230601, People's Republic of China }
	
	\begin{abstract} 
	\vspace{0.5cm}
The paper calculates the helicity-dependent dihadron fragmentation function (DiFF), by extending the dihadron spectator model and examine the single longitudinal spin asymmetry $A^{\sin(2\phi_h)}_{UL}$ from dihadron in semi-inclusive inelastic scattering (SIDIS). This function elucidates the relationship between the longitudinal polarization of the fragmented quark and the transverse momentum of the resulting hadron pairs. A study by the COMPASS collaboration detected a minimal signal in their experimental search for this azimuthal asymmetry in SIDIS. Here, we use the spectator model to calculate the unknown T-odd dihadron fragmentation function $H_1^\perp$. Adopting collinear factorization to describe the data, avoiding the transverse momentum dependent factorization and the associated resummation effects, helping us understand the asymmetry and explaining why the signal is so weak. We involve the approach of transverse momentum dependence in the model calculations, in order to formulate the differential cross sections and the spin asymmetries in terms of the collinear parton distributions and the collinear DiFFs. A transverse momentum factor analysis method was used, in which the transverse momentum of the final hadron pairs was not integrated. The asymmetry of $sin(2\phi_h)$ in COMPASS kinematics was calculated and compared with experimental data.  In addition, predictions for the same asymmetry are also presented for HERMES and the Electron Ion Collider.			
	\end{abstract}
	\maketitle	
	\section{Introduction}
	\label{I}
The study of the dihadron fragmentation functions (DiFFs) describing the probability that a quark hadronizes into two hadrons is of great interest both in theory and in experiment.  The DiFFs were introduced for the first time in Ref.~\cite{Konishi:1979cb}. Their evolution equations have been studied in Ref.~\cite{Vendramin:1981te}. In particular, the authors of Ref.~\cite{Ceccopieri_2007} presented the evolution equations for extended dihadron fragmentation functions explicitly dependent on the invariant mass, $M_h$, of the hadron pair. Then Ref.~\cite{Collins:1994ax} introduced  the transversely polarized fragmentation by using the transversely polarized DiFF, which later lead to the definition of $H_1^\sphericalangle$. Ref.~\cite{Jaffe_1998} also started the whole business on dihadron fragmentation to access the quark transversity distributions. The basis of all possible DiFFs have been given in Ref.~\cite{Bianconi:1999cd}. The authors in Ref.~\cite{Radici:2001na} analysed  the hadron pair system in relative partial waves. In this way, we can make processes of the dihadrons produced clearly. Soon after, the analysis of DiFFs was extended to the subleading twist within a collinear picture~\cite{Bacchetta:2003vn}. It is important to propose  structure functions for the dihadron SIDIS to express the section by using under the framework of structure functions for the dihadron SIDIS~\cite{Gliske:2014wba}. The analysis is complete up to the subleading twist. Researchers began to closely monitor the DiFFs when attempting to extract the chiral-odd transversity distribution. This distribution was initially extracted by considering the Collins effect~\cite{Collins:1992kk} in single hadron SIDIS and back-to-back dihadron production in $e^+e^-$ annihilations Ref.~\cite{Anselmino:2008jk}. Recently, some meaningful results on DiFFs have been given in Refs.~\cite{pitonyak2023number,Cocuzza_2024,Cocuzza:2023vqs}. To comprehend the transversity distribution further, an alternative method involving dihadron Semi-Inclusive Deep Inelastic Scattering (SIDIS) has been recognized, which solely relies on collinear factorization. The chiral-odd DiFF $H_1^\sphericalangle$~\cite{Radici:2001na}, couples with $h_1$ at the leading-twist level. The function $H_1^\sphericalangle$ can be extracted from the production process of two back-to-back hadron pairs in $e^+e^-$ annihilation \cite{Courtoy:2012ry}. The transversity distribution has been determined from dihadron SIDIS and proton proton collision data in previous studies~\cite{Bacchetta:2011ip,Bacchetta:2012ty,Radici:2015mwa,Radici:2016lam,Radici:2018iag}. Model predictions for the DiFFs have been calculated using the spectator model~\cite{Bianconi:1999uc,Bacchetta:2006un,Bacchetta:2008wb,Yang:2019aan}
and the Nambu-Jona-Lasinio (NJL) quark model~\cite{Matevosyan_2014,Matevosyan_2013,Matevosyan_2017,Matevosyan_2018} to estimate the magnitudes of various DiFFs.

The HERMES collaboration~\cite{HERMES:2008mcr} conducted an experiment on azimuthal asymmetry in the dihadron SIDIS process using a transversely polarized proton target. Similarly, the COMPASS collaboration~\cite{Adolph:2012nw,Adolph:2014fjw} also published experimental data on this topic, but with polarized protons and deuterious targets. Additionally, the BELLE collaboration~\cite{HERMES:2008mcr} has measured the azimuthal asymmetry of back-to-back dihadron pair production, leading to the first parameterization of $H_1^\sphericalangle$. The COMPASS collaboration~\cite{Sirtl:2017rhi}, recently gathered experimental data on different azimuthal asymmetries by scattering longitudinally polarized muons off longitudally polarized protons. Theoretical studies have shown that these asymmetries manifest within the collinear factorization framework, with a $\sin(2\phi_h)$ modulation explored in the spectator model~\cite{Luo:2019frz}. The azimuthal angle of the hadron pair system is denoted as $\phi_h$, while $\phi_R$ represents the angle between the lepton plane and dihadron plane. This paper specifically examines the $\sin(2\phi_h)$ modulation. In this work, we adopt collinear factorization to describe the data, and involve the approach of transverse momentum dependence in the model calculations, in order to formulate the differential cross sections and the spin asymmetries in terms of the collinear parton distributions and the collinear DiFFs. TMD expands collinear factorization to include the parton transverse momentum. The COMPASS experiment determined that the $\sin(2\phi_h)$ asymmetry is statistically consistent with 0 within the experimental uncertainty. This study investigates the $\sin(2\phi_h)$ asymmetry based on results from the spectator model for relevant PDFs and DiFFs. Through partial wave expansion, it is found that the only term contributing to this asymmetry is $h_{1L}^\perp H_{UL}^\perp$, where $H_{UL}^\perp$ arises from the interference of two $p$-waves and $h_{1L}$ represent the helicity distribution. Utilizing the spectator model outcomes for the distributions and DiFFs, we assess the $\sin(2\phi_h)$ discrepancy at COMPASS kinematics and compare with the preliminary data from COMPASS.

The paper is structured as follows: Section \ref{II} introduces the theoretical framework of the $\sin(2\phi_h)$ azimuthal asymmetry in dihadron SIDIS with an unpolarized lepton beam scattering off a longitudinally polarized proton target. The application of the spectator model to calculate the T-odd DiFF $H_{1,UL}^\perp$ is discussed in Section \ref{III}. Section \ref{IV} presents the numerical results of the $\sin(2\phi_h)$ azimuthal asymmetry at the COMPASS measurements kinematics and also provides predictions for the EIC. Finally, a summary of the work is presented in Section \ref{V}.
\section{THE $\sin(2\phi_{ h })$ ASYMMETRY OF DIHADRON PRODUCTION IN SIDIS}	
	\label{II}
\begin{figure}[H]
\centering %图片居中
\includegraphics[width=0.5\textwidth]{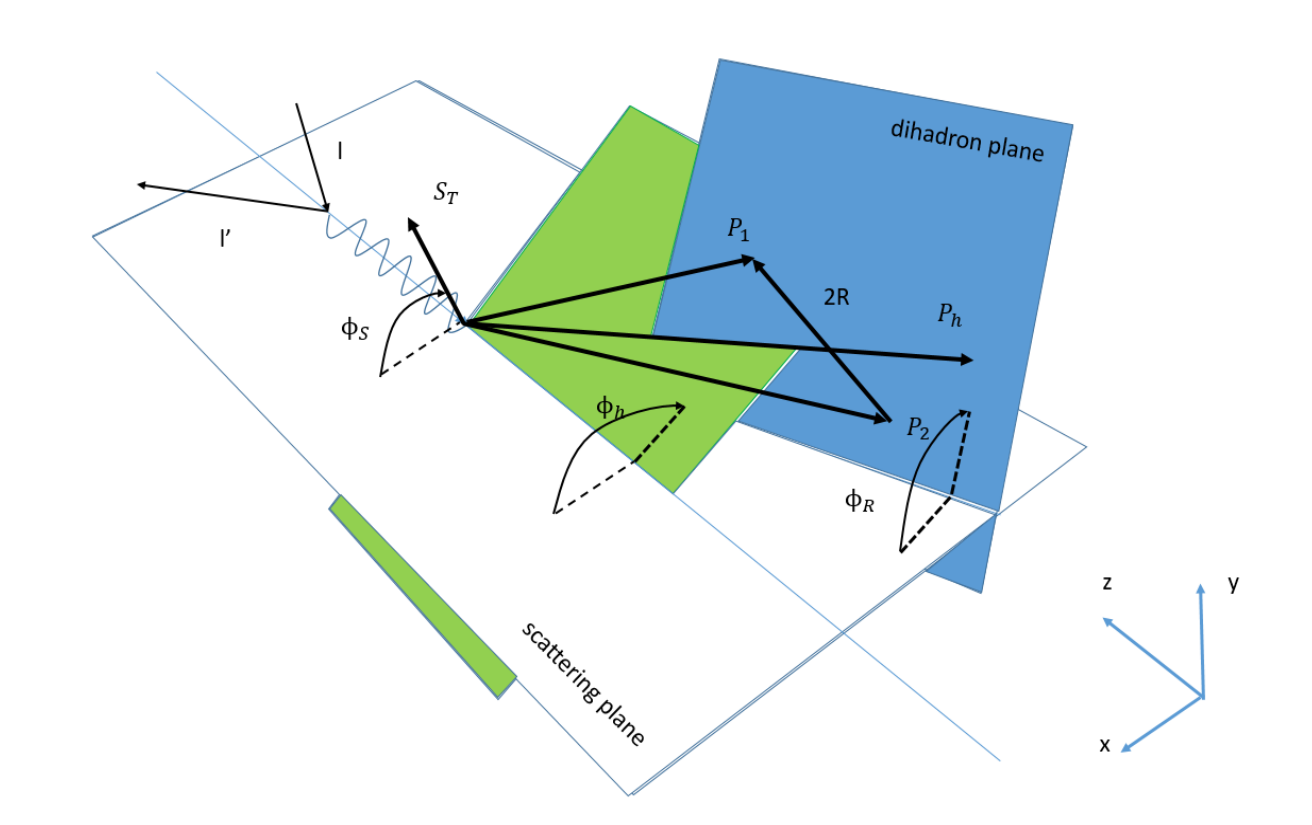} %插入图片，[]中设置图片大小，{}中是图片文件名
\caption{Angle definitions involved in the measurement of the single longitudinal spin asymmetry in SIDIS production of two
	hadrons.} %最终文档中希望显示的图片标题
\label{Fig1} %用于文内引用的标签
\end{figure}

As shown in Fig.~\ref{Fig1}, considering the dihadron fragmentation function $q\rightarrow \pi^+\pi^-X$, where the unpolarized $\mu$ with momentum $\ell$ and a longitudinally polarized beam with mass ${M}$, polarized ${S}$, proton scattering of momentum $P$ by exchanging virtual photons with momentum $q = \ell-\ell^\prime$. In the target, the dynamic quark with momentum $p$ is struck by the photon and the final state quark with momentum $k = p+q$ then fragments into two leading unpolarized hadrons $\pi^+$ and $\pi^-$ with mass $M_1$, $M_2$, and momenta $P_1$, $P_2$. In order to express the differential cross section as well as to calculate the DiFFs, we adopt the following kinematical variables:
	\begin{eqnarray} 
	\begin{aligned}
		&x = \frac { k ^ { + } } { P ^+ } , y = \frac { P \cdot q } { P \cdot \ell } , z = \frac { P _ { h } ^ { - } } { k ^ { - } } = z _ { 1 } + z _ { 2 } ,\\&z _ { i } = \frac { P _ { i } ^ { - } } { k ^ { - } } , Q ^ { 2 } = - q ^ { 2 } , s = ( P + \ell	 ) ^ { 2 } ,\\&P _ { h } = P _ { 1 } + P _ { 2 } , R = \frac { R _ { 1 } - R _ { 2 } } { 2 } , M _ { h } ^ { 2 } = P _ { h } ^ { 2 } .
	\end{aligned}
\end{eqnarray}

Here, we introduce the 4-vector on the frame of the light-cone coordinates as $ a ^ { \mu } = ( a ^ { + } , a ^ { - } , \vec { a } _ { T } ) $, where $ a ^ { \pm } = \frac { a ^ { 0 } \pm a ^ { 3 } } { \sqrt { 2 } }$ and $\vec{a}_T$ is the transverse component of the vector. The light-cone part of the target momentum captured by the initial quark is represented by $x$, $z_i$, and the fragmenting quark is used to represent the light radiation $\pi$ of the hadron. The part of the light-cone of the e fragmenting quark momentum carried by the last pair of hadrons is represented by $z$.  In addition, hadron pairs, the invariant mass, the total momentum, and the relative momentum between them are defined as ${M_h}, P_h$, and $R$, respectively. According to the $\vec{P}_{hT} = 0$, we can make sure the selection of the $\hat{z}$ axis. Hence the momenta $P_h^\mu$, $k_\mu$ and $R_\mu$ can be written as Ref.~\cite{Bacchetta:2006un},

\begin{align}
P_{h}^{\mu}&= \left[ P_{h}^{-}, \frac{M_{h}^{2}}{2P_{h}^{-}}, \vec{0}_{T}\right],\\k^{\mu}&= \left[ \frac{P_{h}^{-}}{z}, \frac{z(k^{2}+ \vec{k}_{T}^{2})}{2P_{h}^{-}}, \vec{k}_{T}\right],\\R^{\mu}&= \bigg[ - \frac{| \vec{R}|P_{h}^{-}}{M_{h}}\cos \theta , \frac{| \vec{R}|M_{h}}{2P_{h}^{-}}\cos \theta \notag,\\&| \vec{R}| \sin \theta \cos \phi _{R},| \vec{R}| \sin \theta \sin \phi _{R}\bigg]\notag,\\&=\left [ - \frac { | \vec{ R } | P _ { h } ^ { - } } { M _ { h } } \cos \theta , \frac { | \vec { R } | M _ { h } } { 2 P _ { h } ^ { - } } \cos \theta , \vec{ R } _ { T } ^ { x } , \vec { R } _ { T } ^ { y }\right ],
\end{align}
where	 	
\begin{align}
	|\vec{R}|&= \sqrt{\frac{M_{h}^{2}}{4}-m_{\pi}^{2}},
\end{align}
here $m_\pi$ is the mass of meson and $\phi_{ R }$ is the angle between the lepton plane and the dihadron plane. It is desired to notice that in order to perform partial-wave expansion, we have reformulated the kinematics in the center of mass frame of the dihadron system. $\theta$ is the center of mass polar angle of the pair with respect to the direction of $P_h$ in the target rest frame~\cite{Bacchetta:2002ux}. There are several useful expression of the scalar products as follows~\cite{Bacchetta:2002ux},
\begin{align}
	&P_{h}\cdot k= \frac{M_{h}^{2}}{2z}+z \frac{k^{2}+ \vec{k}_{T}^{2}}{2},
	\\
	&P_{h}\cdot R=0,
	\\
	&R \cdot k=(\frac{M_{h}}{2z}-z \frac{k^{2}+ \vec{k}_{T}^{2}}{2M_{h}})| \vec{R}| \cos \theta - \vec{k}_{T}\cdot \vec{R}_{T}.
\end{align}

The differential cross section for the SIDIS process with unpolarized muons off a longitudinally polarized nucleon target can be expressed using the TMD factorization approach, denoting $A(y)=1-y+\frac{y^2}{2}$ as shown in Ref.~\cite{Radici:2001na},

\begin{align}\label{eq13}
	\frac { d ^ { 9 } \sigma _ { U U } } { d x d y d z d \phi _ { S } d \phi _ { h } d \phi _ { R } d \cos \theta d \vec { P } _ { h \bot } ^ { 2 } d M _ { h } ^ { 2 } } \notag\\= \frac { \alpha ^ { 2 } } { 2 \pi s x y ^ { 2 } } A ( y ) \sum _ { q } e _ { q } ^ { 2 } \mathcal{L} \left[ f _ { 1 } ^ { q } D _ { 1 , OO } \right] ,
\end{align}
and
\begin{align}
	\frac { d ^ { 9 } \sigma _ { U L } } { d x d y d z d \phi _ { s } d \phi _ { h } d \phi _ { R } d \cos \theta d \vec { P } _ { h \bot } ^ { 2 } d M _ { h } ^ { 2 } } = \notag\\\frac { \alpha ^ { 2 } } { 2 \pi s x y ^ { 2 } } A ( y ) \sum _ { q } e _ { q } ^ { 2 } \{ \sin \theta \sin ( 2\phi _ { h })\times\\\mathcal{L}\left[\frac{2(\vec{p}_T\hat{P}_{h\perp})(\vec{k}_T\hat{P}_{h\perp})-\vec{p}_T\vec{k}_T}{MM_h}\right]h^\perp_{1L}H_{UL}^{\perp}\}\notag.
\end{align}

The azimuthal angles $\phi_{ R }$ and $\phi_{ S }$ represent the angles of $\vec{ R }_T$ and $\vec{S}_T$ with respect to the lepton scattering plane. The angle $\hat{P}_{h\perp}$ is determined by the relationship $\hat{P}_{h\perp} = \vec{ P}_{h\perp}/|\vec{P}_{h\perp}|$. To simplify the labels U and L are used to denote the unpolarized or longitudinally polarized states of the beam or target. The structure functions in Eq.\eqref{eq13} are expressed as weighted convolutions,
\begin{align}
	\mathcal{L} \left[ f \right] = \int d ^ { 2 } \vec { p }_T d ^ { 2 } \vec { k } _ { T } \delta ( \vec { p } _ { T } - \vec { k } _ { T } - \frac { \vec{ P }_ { k _ { \bot }  } } { z } ) \left[ f \right] .
\end{align}

In Eq.~\eqref{eq13}, $f_q^1$ and $D_{1,OO}$ represent the unpolarized PDF and unpolarized DiFF with flavor $q$, respectively. Eq.~\eqref{eqg} introduces $h^\perp_{1L}$ as a twist-2 distribution function associated with the T-odd DiFFs $h_{1L}^\perp$. Both DiFFs play a role in the $\mathrm{sin}(2\phi_{h} )$ azimuthal asymmetry in SIDIS. The expression of the $\sin(2\phi_h)$ asymmetry is as follows,

\begin{widetext}
	\begin{align}\label{eqg}
		A^{\sin(2\phi_h)}_{UL}&=\bigg(\int dxdydz 2M_hdM_hd \cos\theta d^{2}\hat{P}_{h \bot} d^2\vec{k}_T d^2\vec{p}_T(\frac{k_Tp_T\pi}{4})\cdot \delta(p_{T}-k_{T}- \frac{P_{h \bot}}{z})\left[\frac{2(\vec{p}_T\hat{P}_{h\perp})(\vec{k}_T\hat{P}_{h\perp})-\vec{p}_T\vec{k}_T}{MM_h}\right] \notag\\&(4h_{1L}^{\perp u}-h_{1L}^{\perp d})H_{UL}^\perp\bigg)\bigg{/}\bigg({\int dxdydz2M_{h}dM_{h}d \cos \theta d^{2}\hat{P}_{h \bot}d^2\vec{p}_{T}d^2\vec{k}_T\delta(p_{T}-k_{T}- \frac{P_{h \bot}}{z})(4f_{1}^{u}(p_{T}^{2})+f_{1}^{d}(p_{T}^{2}))D_{1,oo}}\bigg).
	\end{align}	
\end{widetext}

	\section{THE MODEL CALCULATION OF $H_{1,UL}^\perp$}
	\label{III}

The TMD DiFFs $D_1$ and $H_1^\perp$ which will appear in the underlying asymmetry are extracted from the quark-quark correlator $\Delta(k; P_h; R)$~\cite{Bacchetta:2003vn},
\begin{align}
	\Delta ( k , P _ { h } , R ) &= \sum \kern -1.3 em \int_X \ \frac { d ^ { 4 } \xi } { ( 2 \pi ) ^ { 4 } } e ^ { i k\cdot \xi } \langle 0 | \psi ( \xi ) | P _ { h } , R ; X \rangle\notag\\& \langle X ; P _ { h } , R | \overline { \psi } ( 0 ) | 0 \rangle | _ { \xi ^ { - } = \vec { \xi }_T  = 0 }\notag\\&=\frac{1}{16\pi}\left\{{D_1\slashed{n}_++H_1^\perp}\frac { \sigma _ { \mu \nu } k _ { T } ^ { \mu } n _ { + } ^ { \nu } } { M _ { h }}\right\} .
\end{align}

Similarly, we need to express the quark quark correlation function formula for the leading order distortion in the center of mass frame, and the relationship between the two correlation functions is as follows:
\begin{align}\label{eq17}
	\Delta ( z , k _ { T } ^ { 2 } , \cos \theta , M _ { h } ^ { 2 } , \phi _ { R } ) = \frac { | \vec { R } | } { 1 6 z M _ { h } } \int d k ^ { + } \Delta ( k , P _ { h } , R ),
\end{align}

By projecting out the usual Dirac structures, we obtain the following decomposition results,
\begin{align}
	4 \pi {\rm T r} \left[ \Delta ( z , k _ { T } , R ) i \sigma ^ { \alpha - } \gamma _ { 5 } \right] = \frac { \varepsilon _ { T } ^ { \alpha \beta } k ^\beta _ { T } } { M _ { h } } H _ { 1 } ^ { \perp } .
\end{align}
where $\gamma^{\alpha-}$ is the negative light-cone Dirac matrix.

The TMD DiFFs $D_1,H_1^\perp$ can be expanded in the relative partial waves of the dihadron system up to the $p$-wave level~\cite{Bacchetta:2002ux}:
\begin{widetext}
	
\begin{align}
	D _ { 1 }(z,k^2_T,\cos \theta,M_h^2)&= D _ { 1 , OO }(z,M_h^2)+ D _ { 1 , O L }(z,M_h^2) \cos \theta + D _ { 1 , L L }(z,M_h^2) \frac { 1 } { 4 } ( 3 \cos ^ { 2 } \theta - 1 )\notag\\& + \cos ( \phi _ { k } - \phi _ { R } ) \sin \theta ( D _ { 1 , OT } + D _ { 1 , L T } \cos \theta ) + \cos ( 2 \phi _ { k } - 2 \phi _ { R } ) \sin ^ { 2 } \theta D _ { 1 , T T }\\ H _ { 1 } ^ { \perp }(z,k^2_T,\cos \theta,M_h^2) &= H _ { 1 , OO }^\perp (z,M_h^2)+ H _ { 1 ,U L } ^ { \perp }(z,M_h^2) \cos \theta + H _ { 1 , L L } ^ { \perp }(z,M_h^2) \frac { 1 } { 4 } ( 3 \cos ^ { 2 } \theta - 1 ) .
\end{align}
\end{widetext}

However, we only expand to the $p$-wave level. here $H_{1,OO}^\perp$ originates from the interference of two $s$-paves, and $H_{1,UL}^\perp$ originates from the interference of two $p$-waves with the different transverse polarizations. $\phi_k$ is the azimuthal angle of quark transverse momentum $\vec{k}_T$ with respect to the lepton scattering plane.

In this work, the above mentioned part of $H_{1,LL}$ does not contribute to the calculation of $\sin(2\phi_h)$, so we only need to calculate the dihidron fragmentation function $H_{1,UL}^\perp$ under the spectator model, here for the sake of simplicity, we will no longer consider the terms related to $\cos\theta$ in the DiFFs expansion, so we will concentrate on calculating $D_{1,OO},H_{ 1,UL}^\perp$. Then we can get something similar function from the Ref.~\cite{Bacchetta:2006un},

\begin{align}\label{eq20}
	\Delta ^ { q } ( k , P _ { h } , R )\notag& = \frac { 1 } { ( 2 \pi ) ^ { 4 } } \frac { ( k + m ) } { ( k ^ { 2 } - m ^ { 2 } ) ^ { 2 } }\\& ( F ^ { s * } e ^ { - \frac { k ^ { 2 } } { \Lambda _ { s } ^ { 2 } } } + F ^ { p * } e ^ { - \frac { k ^ { 2 } } { \Lambda _ { p } ^ { 2 } } } \slashed { R } )\notag\\& ( \slashed{k} -\slashed{ P} _ { h } + M _ { s } ) \times ( F ^ { s } e ^ { - \frac { k ^ { 2 } } { \Lambda _ { s } ^ { 2 } } } + F ^ { p } e ^ { - \frac { k ^ { 2 } } { \Lambda _ { P } ^ { 2 } } } \slashed { R } )\notag\\& (\slashed{k} + m ) \cdot 2 \pi \delta ( ( k - P _ { h } ) ^ { 2 } - M _ { s } ^ { 2 } ) ,
\end{align}

Here $m$ and $M_s$ represent the masses of the fragmented quark and the spectator quark, as denoted by the open circle in Fig.~\ref{Fig2}. $ F^*$,$ F^p$ are the vertices refer to the $s$-wave contribution and $p$-wave contribution~\cite{Bacchetta:2006un}, and as the following forms:

\begin{align}
	&F ^ { s } = f _ { s }\notag \\&F ^ { p } = f _ { \rho } \frac { M _ { h } ^ { 2 } - M _ { \rho } ^ { 2 } - i \Gamma _ { \rho } M _ { \rho } } { ( M _ { h } ^ { 2 } - M _ { \rho } ^ { 2 } ) ^ { 2 } + \Gamma _ { \rho } ^ { 2 } M _ { \rho } ^ { 2 } }+ f _ { \omega } \frac { M _ { h } ^ { 2 } - M _ { \omega } ^ { 2 } - i \Gamma _ { \omega } M _ { \omega } } { ( M _ { h } ^ { 2 } - M _ { \omega } ^ { 2 } ) ^ { 2 } + \Gamma _ { \omega } ^ { 2 } M _ { \omega } ^ { 2 } }  \notag\\&-if_\omega^{\prime}\frac{\sqrt{\lambda(M_\omega^2,M_h^2,m_\pi^2)}\Theta(M_\omega-m_\pi-M_h)}{4\pi\Gamma M_\omega ^2[4M_\omega ^2m_\pi^2+\lambda(M_\omega^2,M_h^2,m_\pi^2)]^{\frac{1}{4}}},
\end{align}
here, $\lambda(M_\omega^2$,$M_h^2,m_\pi^2)=( M _ { \omega } ^ { 2 } - ( M _ { h } + m _ { \pi } ) ^ { 2 } ) ( M _ { \omega } ^ { 2 } - ( M _ { h } - m _ { \pi } ) ^ { 2 } )$, and $\Theta$ denotes the unit step function. The first two terms of $F^p$ can be identified with the contributions of the $\rho$ and the $\omega$ resonances decaying into two pions. The masses and widths of the two resonances are adopted from the PDG(Particle Date Group)~\cite{Fields:2004cb}: $M_\rho=0.776~$GeV, $\Gamma_\rho=0.150~$GeV, $M_\omega=0.783~$GeV.

Putting Eq.~\eqref{eq20} into Eq.~\eqref{eq17}, in this section, we need to calculate $H_{1,UL}^\perp$ with one loop correction. Then according to Feynman rules, we can write the one-loop contribution of the correlation function in Fig.~\ref{Fig1} as:
\begin{widetext}
	\begin{align}
	\Delta_a^q ( z , k _ { T } ^ { 2 } , \cos \theta , M _ { h } ^ { 2 } , \phi _ { R } )&=\frac { C _ { F } \alpha _ { s } } { 3 2 \pi ^ { 2 } ( 1 - z ) P _ { h } ^ { - } } \cdot \frac { | \vec { R } | } { M _ { h } } \cdot \frac { ( \slashed{k} + m ) } { ( k ^ { 2 } - m ^ { 2 } ) ^ { 3 } } ( F ^ { s * } e ^ { - \frac { k ^ { 2 } } { \Lambda _ { 8 } ^ { 2 } } } + F ^ { p * } e ^ { - \frac { k ^ { 2 } } { \Lambda _ { p } ^ { 2 } } }  \slashed{ R } ) (\slashed{ k} - \slashed{P} _ { h } + M _ { s } ) \notag\\&\times ( F ^ { s } e ^ { - \frac { k ^ { 2 } } { \Lambda _ { s } ^ { 2 } } } + F ^ { p } e ^ { - \frac { k ^ { 2 } } { \Lambda _ { p } ^ { 2 } } } \slashed{R} ) (\slashed{ k} + m ) \int \frac { d ^ { 4 } l } { ( 2 \pi ) ^ { 4 } } \frac { \gamma ^ { \mu } ( \slashed{k} - \slashed{l} + m ) \gamma _ { \mu } ( \slashed{k} + m ) } { ( ( k - l ) ^ { 2 } - m ^ { 2 } + i \varepsilon ) ( \varepsilon ^ { 2 } + i \varepsilon ) } ,
\end{align}

\begin{align}
	\Delta _ { b } ^ { q } ( z , k _ { T } ^ { 2 } , \cos \theta , M _ { h } ^ { 2 } , \phi _ { R })& =\frac { C _ { F } \alpha _ { s } } { 3 2 \pi ^ { 2 } ( 1 - z ) P _ { h } ^ { - } } \frac { | \vec{ R } | } { M _ { h } } \frac { ( \slashed{k} + m ) } { ( k ^ { 2 } - m ^ { 2 } ) ^ { 2 } } ( F ^ { s * } e ^ { - \frac { k ^ { 2 } } { \Lambda _ { s } ^ { 2 } } } + F ^ { p * } e ^ { - \frac { k ^ { 2 } } { \Lambda _ { p } ^ { 2 } } } { \slashed{R} } ) (\slashed{ k} -\slashed{ P} _ { h } + M _ { s } )\notag\\&\int \frac { d ^ { 4 } l } { ( 2 \pi ) ^ { 4 } } \frac { \gamma ^ { \mu } ( \slashed{k} - \slashed{P} _ { k } -\slashed{l} + M _ { s } ) ( F ^ { s } e ^ { - \frac { k ^ { 2 } } { \Lambda _ { s } ^ { 2 } } } + F ^ { p } e ^ { - \frac { k ^ { 2 } } { \Lambda _ { p }^2 }  } {\slashed{ R} } ) ( \slashed{k} -\slashed{l}  + m ) \gamma _ { \mu } (\slashed{k} + m ) } { ({ k }-P _ { k } - l ) ^ { 2 } - M _ { s } ^ { 2 } + i \varepsilon ) ( ( k - l ) ^ { 2 } - m ^ { 2 } + i \varepsilon ) ( l ^ { 2 } + i \varepsilon ) } ,
\end{align}

\begin{align}
	\Delta_c^q( z , k _ { T } ^ { 2 } , \cos \theta , M _ { h } ^ { 2 } , \phi _ { R })&=i\frac { C _ { F } \alpha _ { s } } { 3 2 \pi ^ { 2 } ( 1 - z ) P _ { h } ^ { - } } \cdot \frac { | \vec{ R } | } { M _ { h } } \cdot \frac { (\slashed{ k} + m ) } { ( k ^ { 2 } - m ^ { 2 } ) ^ { 2 } } ( F ^ { s * } e ^ { - \frac { k ^ { 2 } } { \Lambda _ { s } ^ { 2 } } } + F ^ { p * } e ^ { - \frac { k ^ { 2 } } { \Lambda _ { p } ^ { 2 } } } \slashed{ R } ) (\slashed{ k} -\slashed{ P} _ { h } + M _ { s } )\notag\\& \times ( F ^ { s } e ^ { - \frac { k ^ { 2 } } { \Lambda _ { s } ^ { 2 } } } + F ^ { p } e ^ { - \frac { k ^ { 2 } } { \Lambda _ { p } ^ { 2 } } } \slashed{ R } ) \int \frac { d ^ { 4 } l } { ( 2 \pi ) ^ { 4 } } \frac { ( \slashed{k} + m ) \gamma ^ { - } ( \slashed{k} - \slashed{l} + m ) } { ( ( k - l ) ^ { 2 } - m ^ { 2 } + i \varepsilon ) ( - l^-  \pm i \varepsilon ) ( l^ { 2 } + i \varepsilon ) } .
\end{align}
\begin{figure}[H] %H为当前位置，!htb为忽略美学标准，htbp为浮动图形
	\centering %图片居中
	\includegraphics[width=0.6\textwidth]{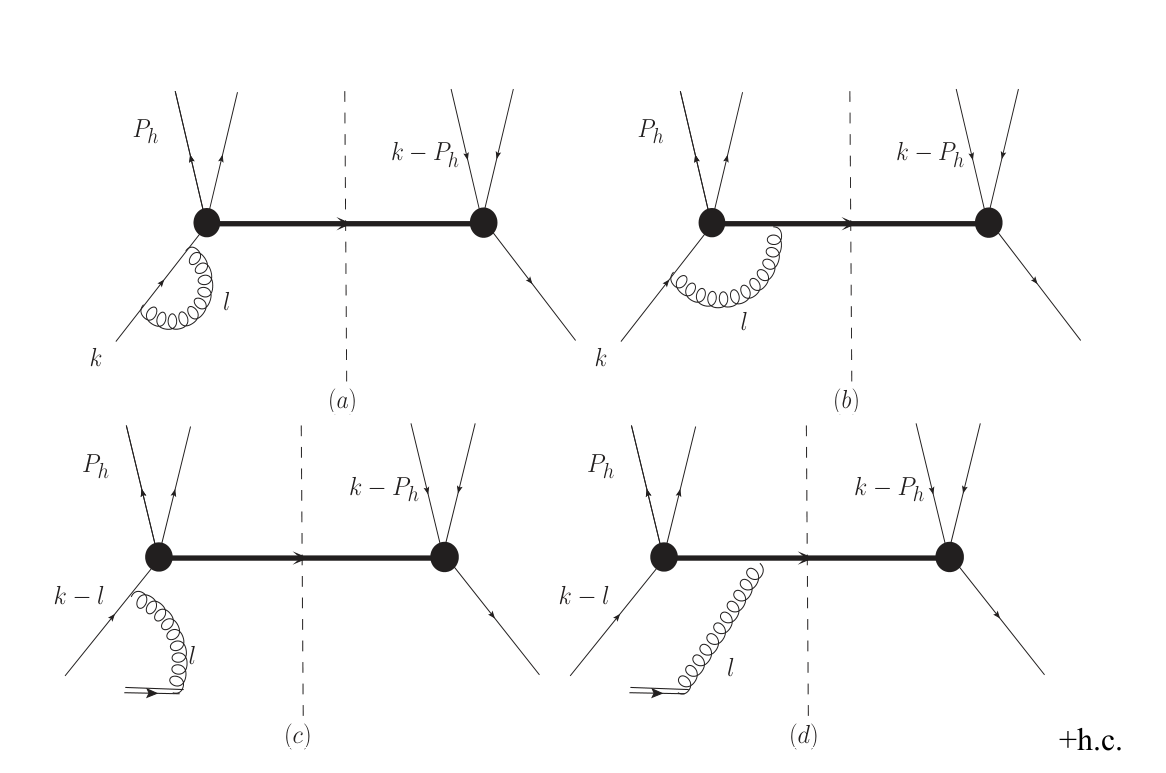} %插入图片，[]中设置图片大小，{}中是图片文件名
	\caption{ One loop order corrections to the fragmentation function of a quark into a meson pair in the spectator model. Where	h.c. represents the hermitian conjugations of these diagrams.} %最终文档中希望显示的图片标题
	\label{Fig2} %用于文内引用的标签
\end{figure}
\begin{align}
	\Delta _ { d } ^ { q } ( z , k _ { T } ^ { 2 } , \cos \theta , M _ { h } ^ { 2 } , \phi _ { R } )& = i \frac { C _ { F } \alpha _ { s } } { 3 2 \pi ^ { 2 } ( 1 - z ) P _ { h } ^ { - } } \cdot \frac { | \vec{ R } | } { M _ { h } } \cdot \frac { ( \slashed{k} + m ) } { k ^ { 2 } - m ^ { 2 } } ( F ^ { s * } e ^ { - \frac { k ^ { 2 } } { \Lambda _ { s } ^ { 2 } } } + F ^ { p * } e ^ { - \frac { k ^ { 2 } } { \Lambda_ { p } ^ { 2 } } } \slashed{ R } ) (\slashed {k} - \slashed{P} _ { h } + m _ { s } )\notag\\&\int \frac { d ^ { 4 } l } { ( 2 \pi ) ^ { 4 } } \frac{\gamma^-(\slashed{k}-\slashed{P}_h-\slashed{l}+m_s)( ( F ^ { s } e ^ { - \frac { k ^ { 2 } } { \Lambda _ { s } ^ { 2 } } } + F ^ { p  } e ^ { - \frac { k ^ { 2 } } { \Lambda_ { p } ^ { 2 } } } \slashed{ R } ) )(\slashed{k}-\slashed{l}+m)}{ ({ k }-P _ { h } - l) ^ { 2 } - M _ { s } ^ { 2 } + i \varepsilon ) ( ( k - l) ^ { 2 } - m ^ { 2 } + i \varepsilon )(-l^-\pm i\varepsilon) ( l ^ { 2 } + i \varepsilon )}.
\end{align}
\end{widetext}
in the above four formulas, the Feynman rules for the optical path propagator $1/(-l^-\pm i\xi)$, as well as the optical path lines and gluon vertices, are used. In the above formula, the Gaussian form factor should depend on the circle momentum $l$. In order to simplify the integral, we follow the choices in Ref.~\cite{Marcel:2021hrv}: discarding the l dependence and only assuming that these Gaussian factors have a $k^2$ dependence, which can give a reasonable final result, similar to the method from Refs.~\cite{Bacchetta:2002tk,Bacchetta:2003xn,Amrath:2005gv}.

In general, there are two sources of $H^\perp_{1,UL}$ for each graph, one is the real part of the circle integral with $|F ^ { p }|^2$ imaginary part is combined, and the imaginary part of the loop integral is combined with $ |F ^ { p }|^2$ real parts are combined, the loop integral real parts are obtained by Feynman parameterization, and the imaginary parts are subject to Cutkosky cutting rule:
\begin{align}
		\frac { 1 } {l ^ { 2 } + i \varepsilon } \rightarrow - 2 \pi i \delta ( l ^ { 2 } ) , \frac { 1 } { ( k - l ) ^ { 2 } + i \varepsilon } \rightarrow - 2 \pi i \delta ( ( k - l ) ^ { 2 } ) .      
\end{align}

Using the above convention, the final result of $H_{1,UL}^\perp$ is obtained:
\begin{widetext}
	\begin{align}\label{eq26}
H_{1,UL}^{\perp a}&= -\frac{1|\vec{ R }|}{4\pi^3(1-z)P_h^-}\cdot |F ^ { p }|^2 e ^ { - \frac { 2 k ^ { 2 } } {\Lambda _ { p } ^ { 2 } } }  ( m ^ { 2 } - { k } ^ { 2 } )(2 m R^-  { k } \cdot R  + R ^ { 2 } \ m\ l^- +R ^ { 2 }\ m\ P_h^- - l^- m _ { s } ),\\H_{1,UL}^{\perp b}&=\frac{|\vec{ R }|}{32\pi^3(1-z)P_h^-}\cdot   |F ^ { p }|^2 e ^ { - \frac { 2 k ^ { 2 } } {\Lambda _ { p } ^ { 2 } } }C_b+\frac { 1 } { 1 6 \pi ^ { 3 } }  \frac { C _ { F } \alpha _ { s } M _ { h } M _ { s } | \vec { R } | } { ( 1 - z ) } \cdot  |F ^ { p }|^2 e ^ { - \frac { 2 k ^ { 2 } } {\Lambda _ { p } ^ { 2 } } }  \frac { k ^ { 2 } } { ( k ^ { 2 } - m ^ { 2 } ) ^ { 2 } }\notag \\&\cdot \ C _ { 2 } ( k ^ { 2 } , M _ { h } ^ { 2 } , 2 k ^ { 2 } + 2 M _ { h } ^ { 2 } - M _ { s } ^ { 2 } , 0 , 0 , M _ { s } ),
\\H_{1,UL}^{\perp c}&=0,
\\H_{1,UL}^{\perp d}&=\frac{|\vec{ R }|}{4\pi^3(1-z)P_h^-}\cdot |F ^ { p }|^2 e ^ { - \frac { 2 k ^ { 2 } } {\Lambda _ { p } ^ { 2 } } }C_d,
	\end{align}
with
	\begin{align}
C_b&=32(m-m_s)(A\ k^--B\ P_h^-)(k\cdot R)^2\notag\\&+R^22(k\cdot P_h(m-m_s)(A\ k^-+B\ P_h^-)+P_h^-\ m_s(m^2-k^2)+I_2\ P_h^-(m_s-m)(k^2-m^2)\notag\\&+8m\ m_s\ A^2(k\cdot R(A\ k^-+B\ P_h^-+P_h^-)(A\ k^-+B\ P_h^-)+I_2\ R^-(m-m_s)(k^2-m^2)-m\ R^-k^2+m^3\ R^-)\notag\\&-2A\ I_2\ R^-\ k\cdot R(m-m_s)(k^2-m^2)+2m\ k\cdot P_h-m\ k^2+m^3)),
\\C_d&=-m\ P_h^-R^2(A\ k^-+B\ P_h^-)-2i\ A\ m\ R^-\ k\cdot R(A\ k^-+B\ P_h^-)\notag\\&+2i\ m\ R^-\ k\cdot R(A\ k^-+B\ P_h^-)+2i\ A\ R^-m_s\ k\cdot R(A\ k^-+B\ P_h^-)+iP_h^-R^2m_s(A\ k^-+B\ P_h^-)\notag\\&+2i\ A\ k^-m\ R^-\ k\cdot R-2i\ A\ k^-R^-m_s -i\ I_2\ m(R^-)^2 (k^2-m^2)+i\ I_2(k^2-m^2),
	\end{align}
\end{widetext}
where $C_2$ is a three-point sing-loop tensor integral, which is defined as:
\begin{align}
	&C _ { 2 } ( p _ { 1 } ^ { 2 } , p _ { 2 } ^ { 2 } , p _ { 3 } ^ { 2 } ; m _ { 1 } ^ { 2 } , m _ { 2 } ^ { 2 } , m _ { 3 } ^ { 2 } )\notag\\& = ( - 2 p _ { 1 } ^ { 2 } B _ { 0 } ( p _ { 1 } ^ { 2 } , m _ { 1 } ^ { 2 } , m _ { 2 } ^ { 2 } ) + ( p _ { 1 } ^ { 2 } + p _ { 2 } ^ { 2 } - p _ { 3 } ^ { 2 } ) \notag\\& \times B _ { 0 } ( p _ { 2 } ^ { 2 } , m _ { 2 } ^ { 2 } , m _ { 3 } ^ { 2 } ) + ( p _ { 1 } ^ { 2 } - p _ { 2 } ^ { 2 } + p _ { 3 } ^ { 2 } ) B _ { 0 } ( p _ { 3 } ^ { 2 } , m _ { 1 } ^ { 2 } , m _ { 3 } ^ { 2 } ) \notag\\&+[ m _ { 1 } ^ { 2 } ( p _ { 1 } ^ { 2 } + p _ { 2 } ^ { 2 } - p _ { 3 } ^ { 2 } )+m_2^2(p_1^2-p_2^2+p_3^2)\notag\\&+p_1^2(-2m_3^2-p_1^2+p_2^2+p_3^2)]C_0(p_1^2,p_2^2,p_3^2;m_1^2,m_2^2,m_3^2),
\end{align}
here $B_0$ is a 2-point one-loop tensor integral, defined as:
\begin{align}
	&B _ { 0 } ( p _ { 1 } ^ { 2 } ; m _ { 1 } ^ { 2 } , m _ { 2 } ^ { 2 } )= \frac { 1 } { i \pi ^ { 2 } } \notag\\\times&\int d ^ { 4 } l \frac { 1 } { ( l ^ { 2 } - m _ { 1 } ^ { 2 } + i \varepsilon ) ( ( l + q _ { 1 } ) ^ { 2 } - m _ { 2 } ^ { 2 } + i \varepsilon ) } ,
\end{align}
in addition, the 3-point one-loop tensor point $C_0$ is defined as:
\begin{align}
	&C _ { 0 } ( p _ { 1 } ^ { 2 } , p _ { 2 } ^ { 2 } , p _ { 3 } ^ { 2 } ; m _ { 1 } ^ { 2 } , m _ { 2 } ^ { 2 } , m _ { 3 } ^ { 2 } ) =\frac { 1 } { i \pi ^ { 2 } } \int d ^ { 4 } l
	 \frac { 1 } { ( l ^ { 2 } - m _ { 1 } ^ { 2 } + i \varepsilon )}\notag\\& \times\frac{1}{( ( l + q _ { 1 } ) ^ { 2 } - m _ { 2 } ^ { 2 } + i \varepsilon )}\cdot\frac{1}{ ( ( l + q _ { 2 } ) ^ { 2 } - m _ { 3 } ^ { 2 } + i \varepsilon ) } ,
\end{align}
where $ q _ { n } \equiv \sum _ { i = 1 } ^ { n } p _ { i }$ and $q_0=0$. The coefficients A and B denote the following functions:

\begin{align}
	 A &= \frac { I _ { 1 } } { \lambda ( k , M _ { h } , M _ { s } ) } [ 2 k ^ { 2 } ( k ^ { 2 } - M _ { s } ^ { 2 } - M _ { h } ^ { 2 } ) \frac { I _ { 2 } } { \pi }\notag\\& + ( k ^ { 2 } + M _ { h } ^ { 2 } - M _ { s } ^ { 2 } ) ] ,\\ B &= - \frac { 2 k ^ { 2 } } { \lambda ( k , M _ { h } ^ { 2 } , M _ { s } ^ { 2 } ) } I _ { 1 } \left(1 + \frac { k ^ { 2 } + M _ { s } ^ { 2 } - M _ { h } ^ { 2 } } { \pi } I _ { 2 } \right),
\end{align}
which originate from the decomposition of the following integral Ref.~\cite{Lu:2015wja},
\begin{align}
	\int d ^ { 4 } l \frac { \rho ^ { \mu } \delta ( l^ { 2 } ) \delta \left[ ( k - l ) ^ { 2 } - m ^ { 2 } \right] } { ( k - P _ { h } - l ) ^ { 2 } - M _ { s } ^ { 2 } } = A k ^ { \mu } + B P _ { h } ^ { \mu } ,
\end{align} 
The functions $I_i$ represent the results of the following integrals:
\begin{align}
	& I _ { 1 } = \int d ^ { 4 } l \delta ( l ^ { 2 } ) \delta \left[ ( k - l ) ^ { 2 } - m ^ { 2 } \right] = \frac { \pi } { 2 k ^ { 2 } } ( k ^ { 2 } - m ^ { 2 } ) ,\\&I _ { 2 } = \int d ^ { 4 } l \frac { \delta ( l^ { 2 } ) \delta \left[ ( k - l ) ^ { 2 } - m ^ { 2 } \right] } { ( k -l - P _ { h } ) ^ { 2 } - M _ { s } ^ { 2 } } = \notag\\&\frac { \pi } { 2 \sqrt { \lambda ( k , M _ { h } , M _ { s } ) } } \ln \bigg( 1 - \frac { 2 \sqrt { \lambda ( k , M _ { h } , M _ { s } ) } } { k ^ { 2 } - M _ { h } ^ { 2 } + M _ { s } ^ { 2 } + \sqrt { \lambda ( k , M _ { h } , M _ { s } ) } } \bigg).
\end{align}

\section{NUMERICAL RESULTS}
	\label{IV}
In the frame of spectator model, we choose the values for the parameters $\alpha_{s,p}$, $\beta_{s,p}$, $\gamma_{s,p}$, $m$, $m_s$ from Ref.~\cite{Bacchetta:2006un}, where the model parameters were tuned to the output of PYTHIA event generator adopted for HERMES~\cite{Sjostrand:2000wi}:
\begin{align}
	&\alpha _ { s } = 2 . 6 0 ~\mathrm {G e V}  ~~\beta _ { s } = - 0 . 7 5 1  ~~\gamma _ { s } = - 0 . 1 9 3 ,\\&\alpha _ { p } = 7 . 0 7 ~\mathrm {G e V}  ~~\beta _ { p } = - 0 . 0 3 8  ~~\gamma _ { p } = - 0 . 0 8 5 ,\\& f_s=1197~\mathrm{GeV}^{-1}  ~~f _ { \rho } = 9 3 . 5 ~~ f _ { \omega } = 0 . 6 3 ,\\ &f _ { \omega } ^ { \prime } = 7 5. 2 ~~M _ { s } = 2 . 9 7~M _ { h } ~~ m = 0 . 0 ~\mathrm{GeV}.
\end{align}

Following the convention in Ref.~\cite{Bacchetta:2006un}, we set the quark mass $m$ to 0 GeV. Additionally, we approximate the strong coupling constant $\alpha_{ s }\approx0.3$.

\begin{figure*}  %H为当前位置，!htb为忽略美学标准，htbp为浮动图形
	\centering %图片居中
	\includegraphics[width=0.4\textwidth]{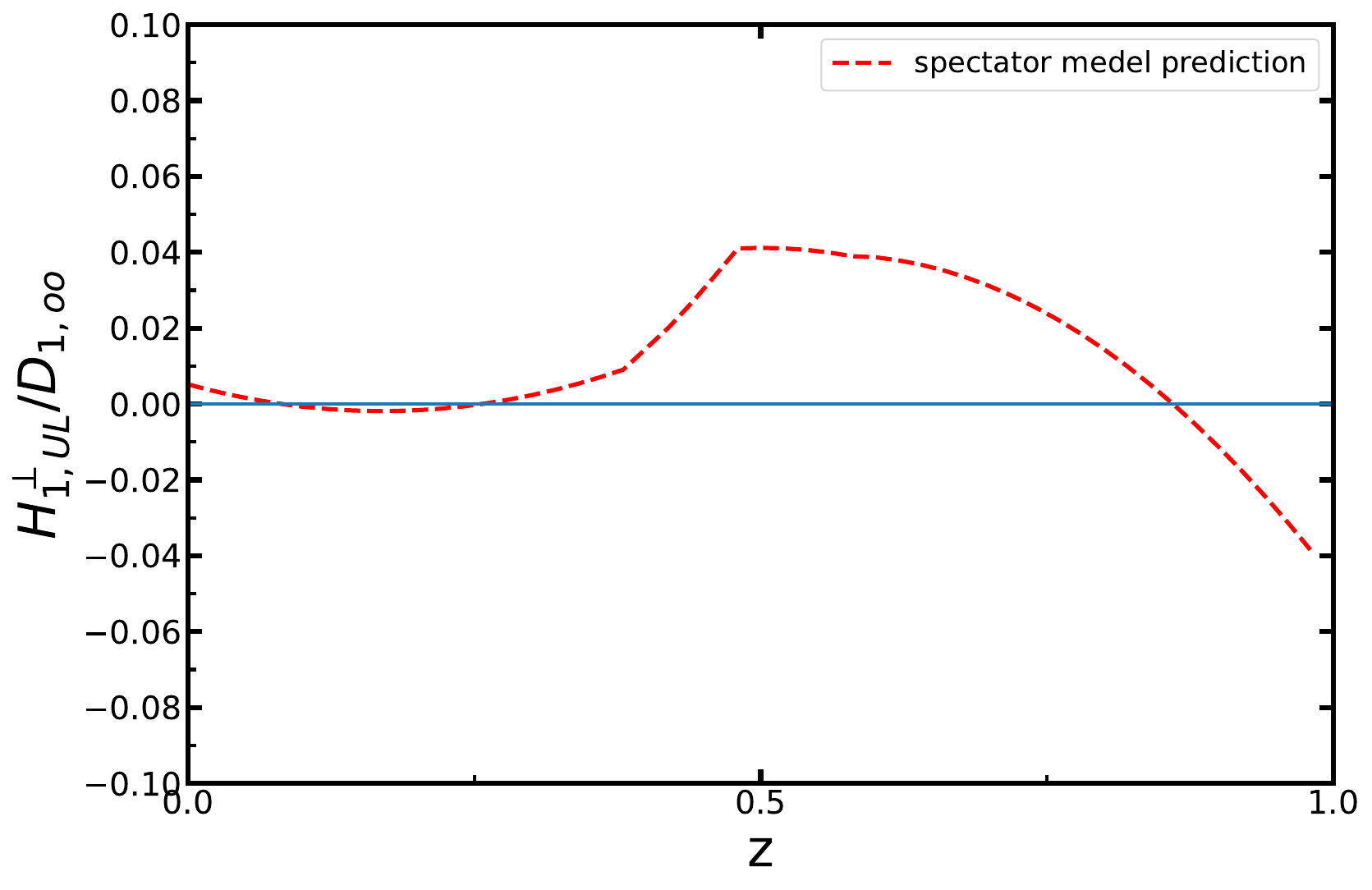}
	\includegraphics[width=0.4\textwidth]{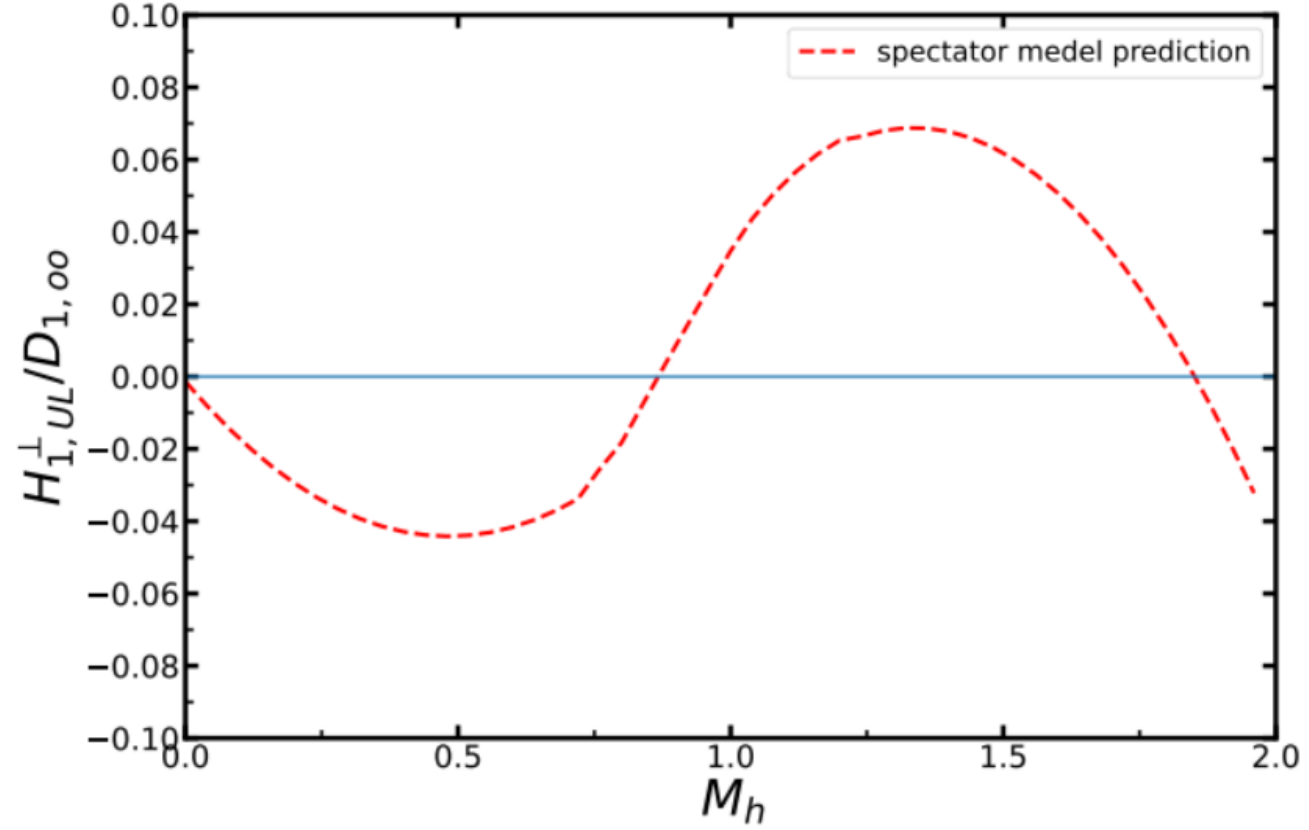} %插入图片，[]中设置图片大小，{}中是图片文件名 
	\caption{The DiFF $H_{1,UL}^\perp$ as functions of $z$ (left panel) and $M_h$ (right panel) in the spectator model, normalized by the unpolarized DiFF $D_{1,OO}$.} %最终文档中希望显示的图片标题
	\label{Fig3} %用于文内引用的标签
\end{figure*}

The ratio between $H_{1,UL}^\perp$and $D_{1,OO}$ is analyzed as a function of $z$ or $M_h$, integrated over the region $\mathrm{0.3~ GeV < M_h < 1.6~GeV }$ or $\mathrm{0.2 < z < 0.9}$ in the left panel and right panel of Fig.~\ref{Fig3} respectively. When compared to the unpolarized DiFF $D_{1,OO}$ Ref.\cite{Bacchetta:2006un}, In left image shows a peak around $z=0.5~\mathrm{GeV}$, while the right image shows a minimum around $M_h=0.5~\mathrm{GeV}$ and maximum values around $M_h=1.3~\mathrm{GeV}$, with a magnitude of approximately $10^{-2}$. The trend displays an initial decrease followed by an increase.

The numerical results of the azimuth asymmetry of $\sin(2\phi_h)$ in the SIDIS process generated by dihadron are given below, with a non-polarized $\mu$ and a longitudinally polarized nucleon target scattering. According to the isospin symmetry, the fragmentation correlation function is found to be important for the processes $\mu \rightarrow\pi^+\pi^-X$, $\bar{d} \rightarrow \pi^+\pi^-X$, $d \rightarrow \pi^-\pi^+X$ and $\bar{u} \rightarrow \pi^-\pi^+X$ are the same. So by transforming the sign of $\vec{R}$, or by doing the $\theta \rightarrow \pi - \theta$ and $\varphi \rightarrow \varphi + \pi$ transformations equivalently, the linearly dependent dihadron fragmentation function from the $d \rightarrow \pi^-\pi^+X$ and $\bar{\mu} \rightarrow \pi^-\pi^+X$ processes $H_{1,UL}^\perp$ has an additional minus sign relative to the one from the $\bar{\mu} \rightarrow \pi^+\pi^-X$ process,
$\sin(2\phi_h)$ asymmetry can be adopted as follows:
\begin{widetext}

	\begin{align}
		A^{\sin(2\phi_h)}_{UL}(x)&=\bigg(\int dydz 2M_hdM_hd \cos\theta d^{2}\hat{P}_{h \bot} d^2\vec{k}_T d^2\vec{p}_T\left(\frac{k_Tp_T\pi}{4}\right)\cdot \delta\left(p_{T}-k_{T}- \frac{P_{h \bot}}{z}\right)\left[\frac{2(\vec{p}_T\hat{P}_{h\perp})(\vec{k}_T\hat{P}_{h\perp})-\vec{p}_T\vec{k}_T}{MM_h}\right]\notag \\\times&(4h_{1L}^{\perp u}-h_{1L}^{\perp d})H_{UL}^\perp\bigg) \bigg{/}\bigg({\int dydz2M_{h}dM_{h}d \cos \theta d^{2}\hat{P}_{h \bot}d^2\vec{p}_{T}d^2\vec{k}_T\cdot\delta\left(p_{T}-k_{T}- \frac{P_{h \bot}}{z}\right)(4f_{1}^{u}(p_{T}^{2})+f_{1}^{d}(p_{T}^{2}))D_{1,oo}}\bigg),
	\end{align}
	\begin{align}
		A^{\sin(2\phi_h)}_{UL}(z)&=\bigg(\int dxdy 2M_hdM_hd \cos\theta d^{2}\hat{P}_{h \bot} d^2\vec{k}_T d^2\vec{p}_T\left(\frac{k_Tp_T\pi}{4}\right)\cdot \delta\left(p_{T}-k_{T}- \frac{P_{h \bot}}{z}\right)\left[\frac{2(\vec{p}_T\hat{P}_{h\perp})(\vec{k}_T\hat{P}_{h\perp})-\vec{p}_T\vec{k}_T}{MM_h}\right]\notag \\\times&(4h_{1L}^{\perp u}-h_{1L}^{\perp d})H_{UL}^\perp\bigg)\bigg{/}\bigg({\int dxdy2M_{h}dM_{h}d \cos \theta d^{2}\hat{P}_{h \bot}d^2\vec{p}_{T}d^2\vec{k}_T\cdot\delta\left(p_{T}-k_{T}- \frac{P_{h \bot}}{z}\right)(4f_{1}^{u}(p_{T}^{2})+f_{1}^{d}(p_{T}^{2}))D_{1,oo}}\bigg),
	\end{align}
	\begin{align}
		A^{\sin(2\phi_h)}_{UL}(M_h)&=\bigg(\int dxdydz 2M_hd \cos\theta d^{2}\hat{P}_{h \bot} d^2\vec{k}_T d^2\vec{p}_T\left(\frac{k_Tp_T\pi}{4}\right)\cdot \delta\left(p_{T}-k_{T}- \frac{P_{h \bot}}{z}\right)\left[\frac{2(\vec{p}_T\hat{P}_{h\perp})(\vec{k}_T\hat{P}_{h\perp})-\vec{p}_T\vec{k}_T}{MM_h}\right]\notag\\\times&(4h_{1L}^{\perp u}-h_{1L}^{\perp d})H_{UL}^\perp\bigg)\bigg{/}\bigg({\int dxdydz2M_{h}d \cos \theta d^{2}\hat{P}_{h \bot}d^2\vec{p}_{T}d^2\vec{k}_T\cdot\delta\left(p_{T}-k_{T}- \frac{P_{h \bot}}{z}\right)(4f_{1}^{u}(p_{T}^{2})+f_{1}^{d}(p_{T}^{2}))D_{1,oo}}\bigg),
	\end{align}	
\end{widetext}
where the TMD DiFF $D_{1,OO}$ has been worked out and listed as Ref.~\cite{Luo:2019frz},
\begin{align}
	&D _ { 1 , O O } ( z , \vec { k } _ { T } ^ { 2 } , M _ { h } ) = \frac { 4 \pi | \vec { R } | } { 2 5 6 \pi ^ { 3 } M _ { h } z ( 1 - z ) ( k ^ { 2 } - m ^ { 2 } ) ^ { 2 } } \notag\\&\Bigg\{ 4 | F ^ { s } | ^ { 2 } e ^ { - \frac { 2 k ^ { 2 } } { \Lambda _ { s } ^ { 2 } } } ( z k ^ { 2 } - M _ { h } ^ { 2 } - m ^ { 2 } z + m ^ { 2 } + 2 m M _ { s } + M _ { s })  - \notag\\&4 | F ^ { p } | ^ { 2 } e ^ { - \frac { 2 k ^ { 2 } } { \Lambda _ { p } ^ { 2 } } } | \vec{ R } | ^ { 2 } ( - z k ^ { 2 } + M _ { h } ^ { 2 } + m ^ { 2 } ( z - 1 ) + 2 m M _ { s } - M _ { s } ^ { 2 } )\notag\\&+ \frac { 4 } { 3 } | F ^ { p } | ^ { 2 } e ^ { - \frac { 2 k ^ { 2 } } { \Lambda _ { P } ^ { 2 } } } | \vec{ R } | ^ { 2 }\bigg[ 4 \bigg( \frac { M _ { h } } { 2 z } - z \frac { k ^ { 2 } + \vec { k } _ { T } ^ { 2 } } { 2 M _ { h } } \bigg)\notag\\& + 2 z \frac { k ^ { 2 } - m ^ { 2 } } { M _ { h } } \bigg( \frac { M _ { h } } { 2 z } - z \frac { k ^ { 2 } + \vec{ k } _ { T } ^ { 2 } } { 2 M _ { h } }\bigg )\bigg]\Bigg\}.
\end{align}

\begin{figure*} %H为当前位置，!htb为忽略美学标准，htbp为浮动图形
	\centering %图片居中
    \subfigure[]{\includegraphics[width=0.4\textwidth]{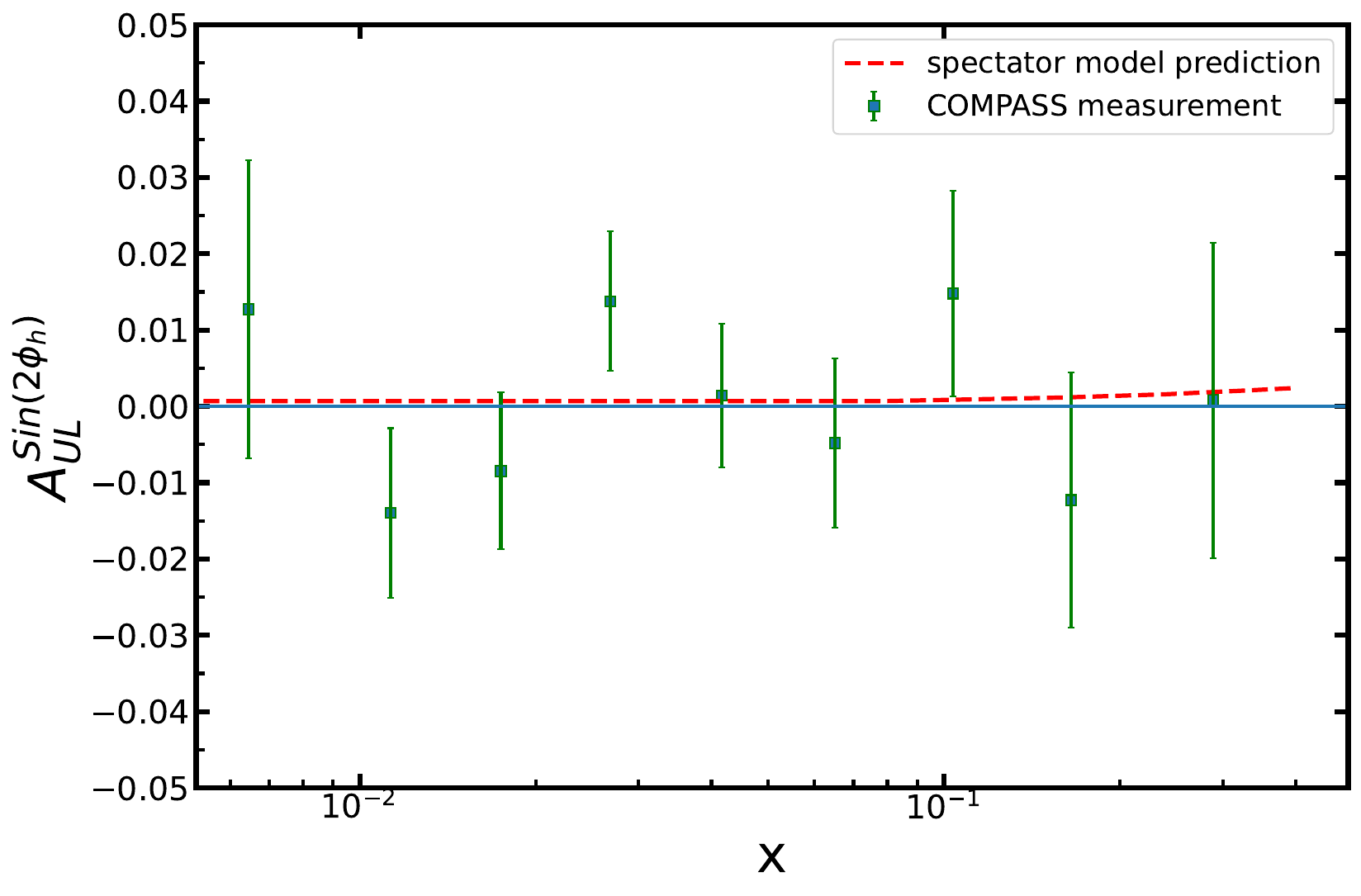}}
	\subfigure[]{\includegraphics[width=0.4\textwidth]{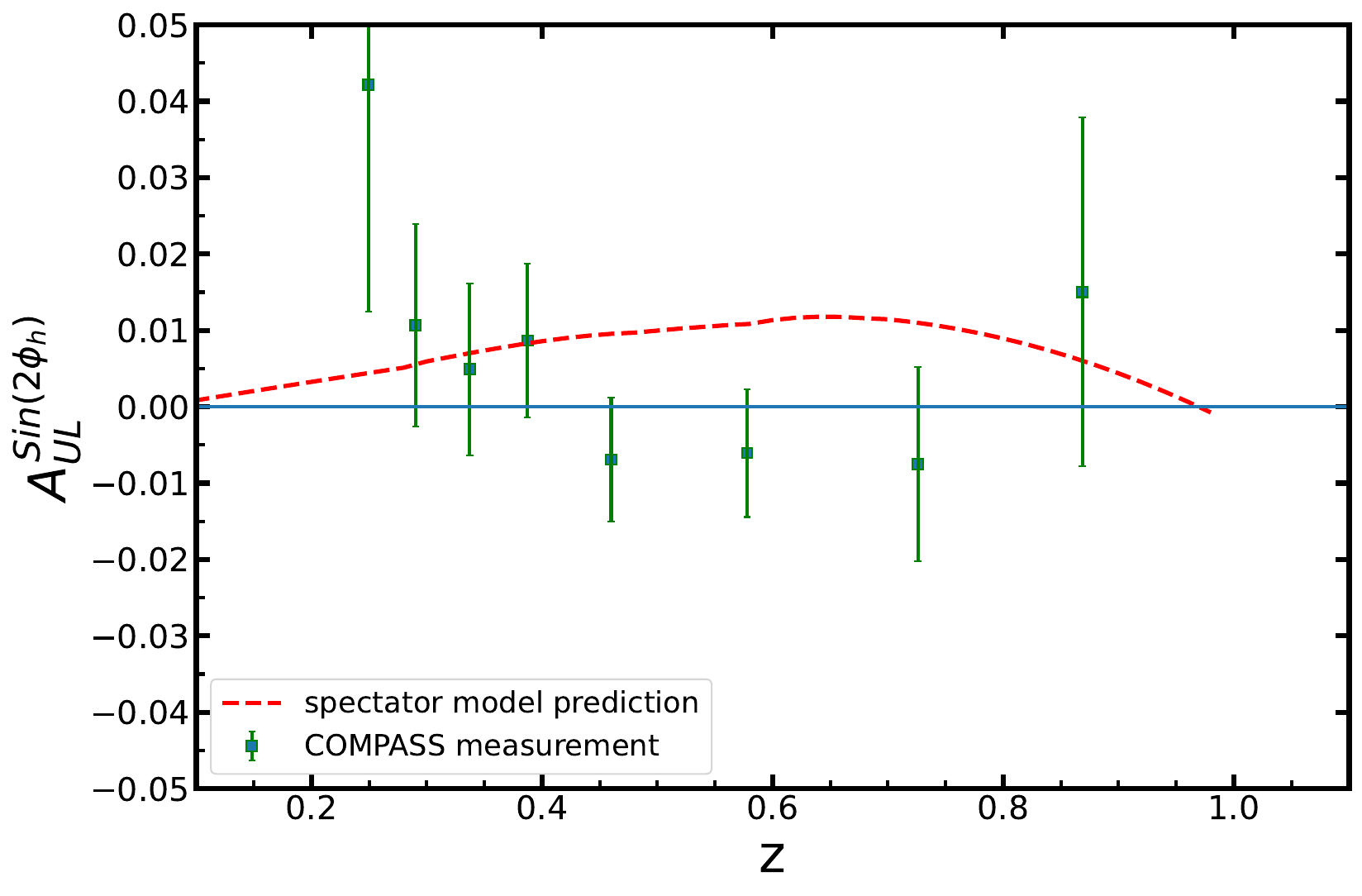}} %插入图片，[]中设置图片大小，{}中是图片文件名
	\subfigure[]{\includegraphics[width=0.4\textwidth]{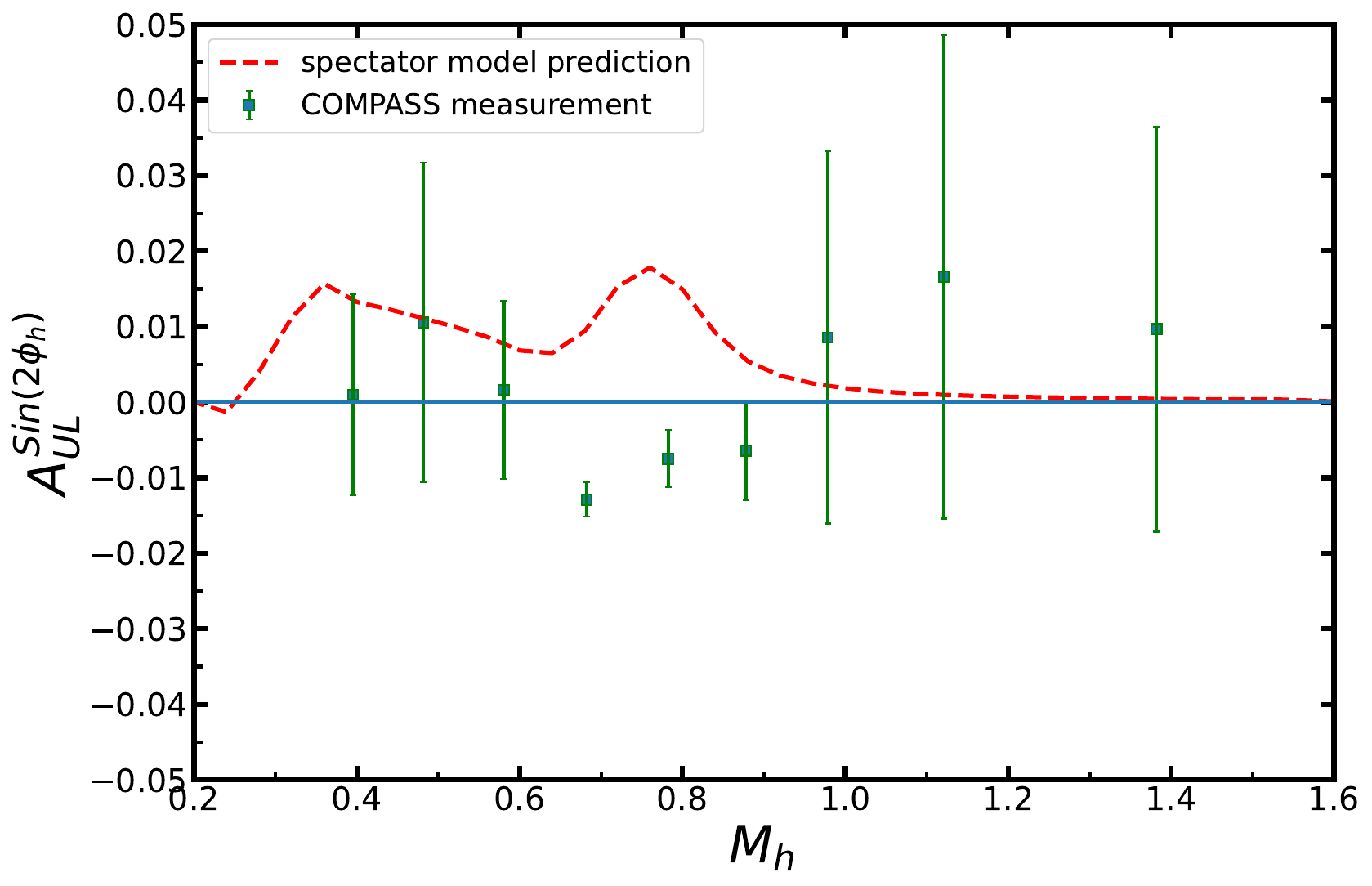}}
	\caption{The $\sin(2\phi_h)$ azimuthal asymmetry in the SIDIS process of unpolarized muons off longitudinally polarized
		nucleon target as a functions of $x$ (figure a), $z$ (figure b) and $M_h$ (figure c) at COMPASS. The full circles with error bars show the preliminary COMPASS data for comparison. The dashed curves denote the model prediction.} %最终文档中希望显示的图片标题
	\label{Fig4} %用于文内引用的标签
\end{figure*}

As for the twist-2 PDFs $f_1$ and $h_{1L}$, we adopt the same spectator model results~\cite{Bacchetta:2008af} for uniformity. To perform numerical calculation for the $\sin(2\phi_h)$ asymmetry in dihadron SIDIS at the COMPASS kinematics, we adopt the following kinematical cuts~\cite{sirtl2017azimuthal},
\begin{align}
	&\mathrm{\sqrt { s } = 1 7 . 4 ~G e V}~~~~~0 . 0 0 3 < x < 0 . 4~,~~~~~\notag\\&0 . 1 < y < 0 . 9~~~~~ 0 . 2 < z < 0 . 9~~~~~Q ^ { 2 } > 1~\mathrm{G e V} ^ { 2 },\notag\\&W > 5 ~\mathrm{G e V}~~~~~\mathrm{ 0 . 3~ G e V} < M _ { h } < 1 . 6 ~\mathrm{G e V},
\end{align}
where $W$ is the invariant mass of photon-nucleon system with $W^2=(P+q)^2\approx\frac{1-x}{x}Q^2$, our main results are plot in Fig.~\ref{Fig4}, showing the predictions for the $\sin(2\phi_h)$ azimuthal asymmetry in the SIDIS process with unpolarized muons off
longitudinally polarized nucleon target, as shown in Fig.~\ref{Fig1}. The $x$-, $z$- and $M_h$-dependent asymmetries are depicted in Fig.~\ref{Fig4} (a), (b), (c). The model predictions are represented by solid lines, while the preliminary COMPASS data is shown with full circles and error bars for comparison. The measured asymmetry values are comparable in magnitude to the test results, and the distributions of $x$ and $z$ appear relatively flat, warranting further consideration. Specifically focusing on the asymmetry distribution with respect to $M_h$, Figure c illustrates theoretical calculations aligning closely with experimental trends, featuring peaks around $M_h=0.38$ and $M_h=0.75$ before approaching 0.35. It is important to note that the model results do not take into account QCD evolution effects,  therefore only providing a rough prediction of the COMPASS preliminary data. Later works that incorporate QCD evolution are likely to yield more reliable predictions.

\begin{figure*} %H为当前位置，!htb为忽略美学标准，htbp为浮动图形
	\centering %图片居中
	\subfigure[]{\includegraphics[width=0.4\textwidth]{EICx.pdf}}
	\subfigure[]{\includegraphics[width=0.4\textwidth]{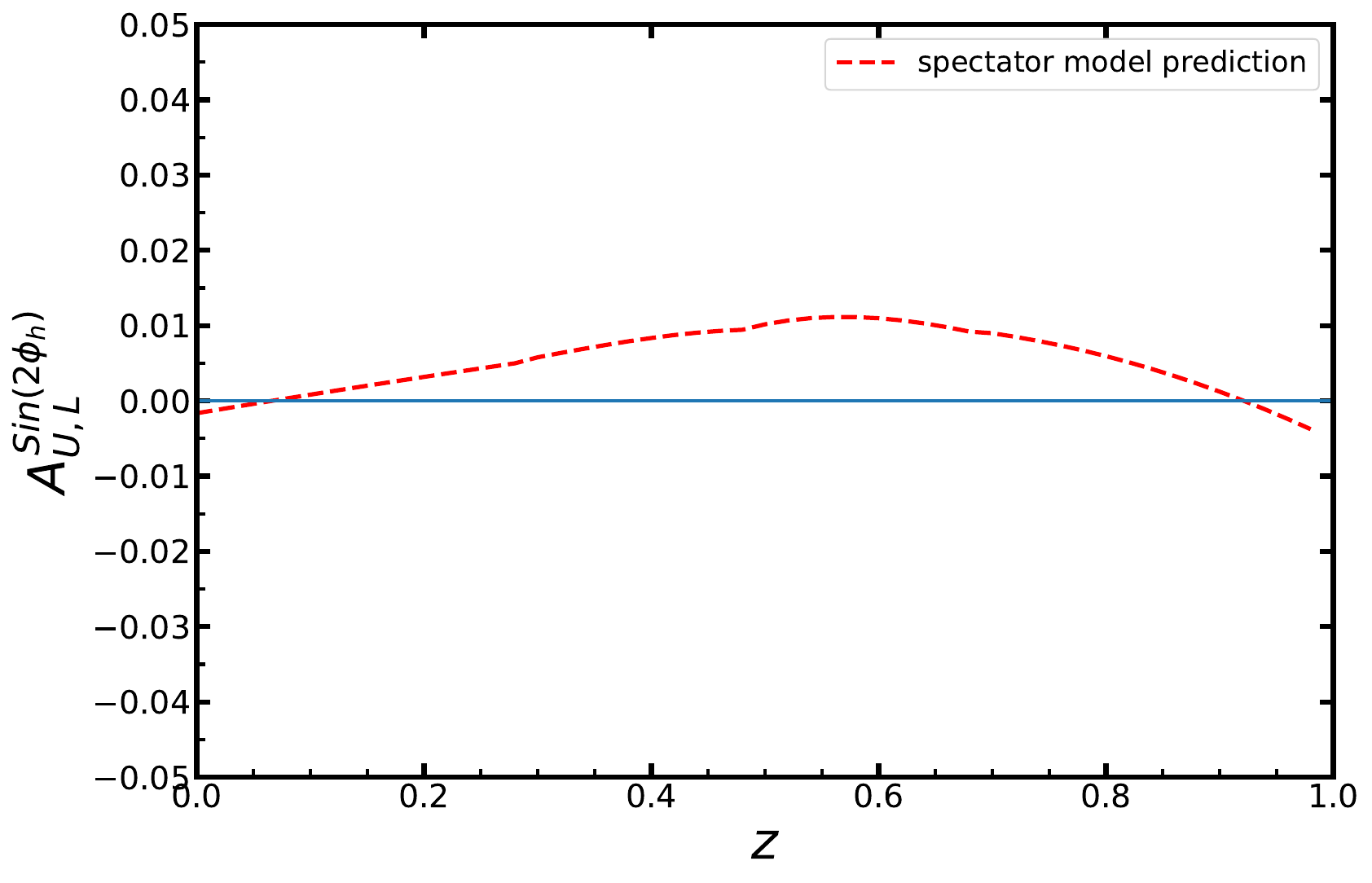}}
	\subfigure[]{\includegraphics[width=0.4\textwidth]{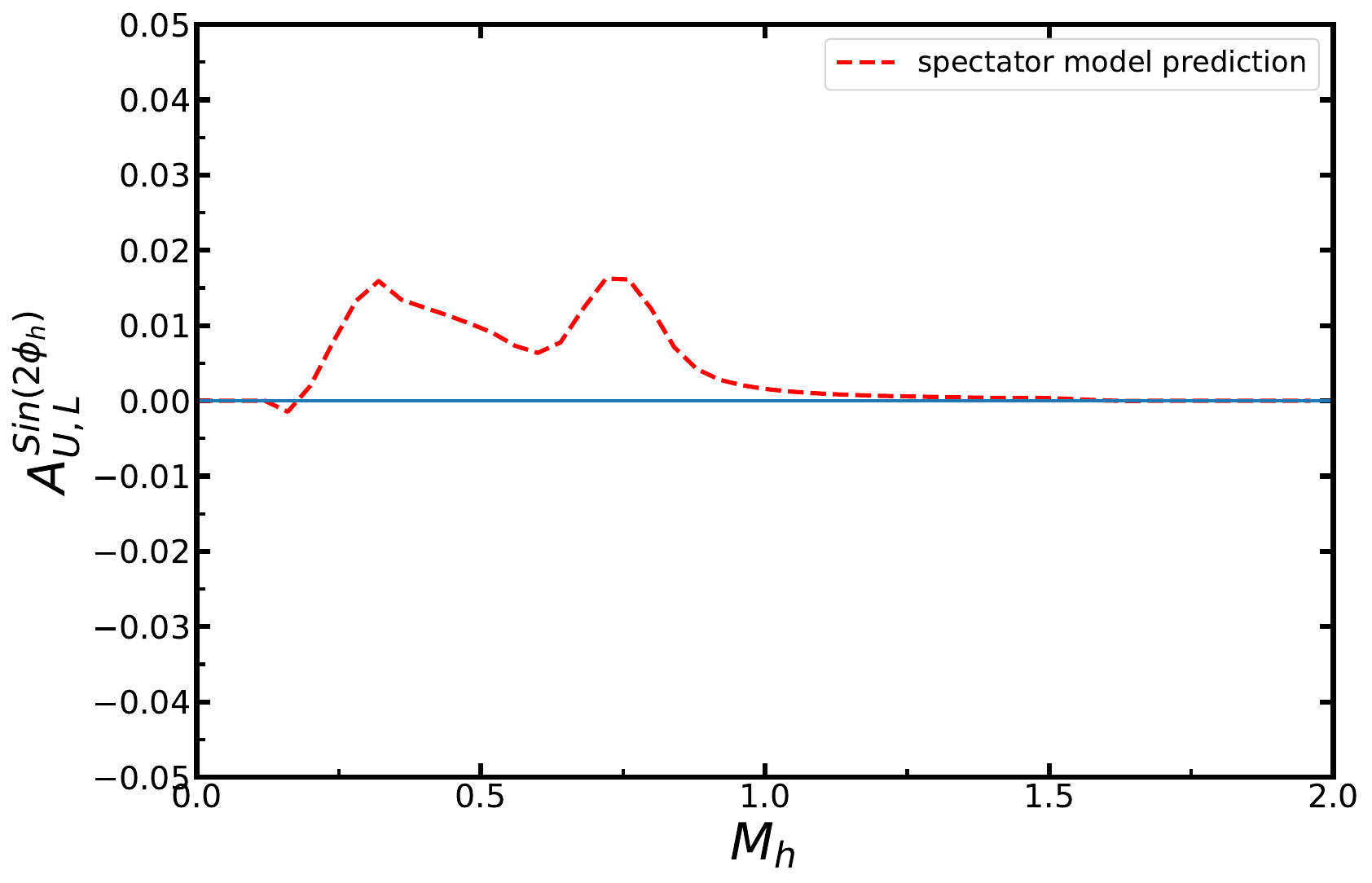}}	
%插入图片，[]中设置图片大小，{}中是图片文件名
	\caption{The $\sin(2\phi_h)$ azimuthal asymmetry in the SIDIS process of unpolarized muons off longitudinally polarized nucleon target as a functions of $x$ (figure a), $z$ (figure b) and $M_h$ (figure c) at the EIC $(\sqrt{s} = 45~\mathrm{GeV})$. The dashed curves denote the model prediction} %最终文档中希望显示的图片标题
	\label{Fig5} %用于文内引用的标签
\end{figure*}

In addition, we also predict the $\sin(2\phi_h)$ asymmetry in the double longitudinally polarized SIDIS at the future EIC. Such a facility would be ideal for studying this observable, as shown in Fig.~\ref{Fig5}. We will use the EIC kinematical cuts outlined in Ref.~\cite{Accardi:2012qut}: 
\begin{align}
		&\mathrm{\sqrt { s } = 45.0 ~G e V}~~~~0 . 0 0 3 < x < 0 . 4~~~~0 . 1 < y < 0 . 9,\notag\\&0 . 2 < z < 0 . 9~~~~~
		Q ^ { 2 } > 1~\mathrm{G e V} ^ { 2 }~~~~~W > 5 ~\mathrm{G e V},\notag\\&\mathrm{ 0 . 3~ G e V} < M _ { h } < 1 . 6 ~\mathrm{G e V} .
\end{align}

\section{CONCLUSION}
	\label{V}
In this study, we examined the single spin asymmetry involving a $\sin(2\phi_h)$ modulation in dihadron production within the framework of SIDIS. Utilizing the spectator model outcome for $D_{1,OO}$, we calculated the T-odd DiFF $H_ {1,UL}^\perp$ by analyzing both the real and imaginary loop contributions. Utilizing the partial wave expansion, it was determined that $H_ {1,UL}^\perp$ arises from the interference of $s$ and $p$-waves. Through the analysis of numerical results from DiFFs and PDFs, we offer a prediction for the $\sin(2\phi_h)$ asymmetry and compare it with measurements from COMPASS.  Our result yields a good description of the vanished COMPASS data. At the HERMES and EIC kinematics we also obtain a very small asymmetry.

\section{Acknowledgments}
\label{VI}
This work is partly supported by the National Natural Science Foundation of China under Grants No. 12205002,
partly supported by the the Natural Science Foundation of Anhui Province (2108085MA20,2208085MA10), 
and partly supported by the key Research Foundation of Education Ministry of Anhui Province of China (KJ2021A0061).
	
\bibliography{art}
\end{document}